# A General SIMD-based Approach to Accelerating Compression Algorithms

WAYNE XIN ZHAO, Renmin University of China
XUDONG ZHANG, Yahoo! China
DANIEL LEMIRE, Université du Québec
DONGDONG SHAN, Alibaba Group
JIAN-YUN NIE, Université de Montréal
HONGFEI YAN, Peking University
JI-RONG WEN, Renmin University of China

Compression algorithms are important for data oriented tasks, especially in the era of "Big Data". Modern processors equipped with powerful SIMD instruction sets, provide us an opportunity for achieving better compression performance. Previous research has shown that SIMD-based optimizations can multiply decoding speeds. Following these pioneering studies, we propose a general approach to accelerate compression algorithms. By instantiating the approach, we have developed several novel integer compression algorithms, called Group-Simple, Group-Scheme, Group-AFOR, and Group-PFD, and implemented their corresponding vectorized versions. We evaluate the proposed algorithms on two public TREC datasets, a Wikipedia dataset and a Twitter dataset. With competitive compression ratios and encoding speeds, our SIMD-based algorithms outperform state-of-the-art non-vectorized algorithms with respect to decoding speeds.

## 1. INTRODUCTION

In recent years, we have witnessed an explosive growth of Web data. The overwhelming data raises compelling computational challenges to Web search engines. Although nowadays CPUs have powerful computational ability, the performance of Web search engines is largely inhibited by slow disk accesses, and the bandwidth of data transferred from disk to main memory becomes the limiting factor for the efficiency.

For search engines, the performance of the primary structure, i.e., the inverted index, is a priority. Various techniques have been shown to be effective to improve the performance of inverted indexes, especially index compression [Navarro et al. 2000]. Compression algorithms can reduce the space of posting lists, and therefore reduce the transfer of data from disk to memory [Manning et al. 2008, p. 85; Zhang et al. 2008]. To improve the efficiency of query evaluation, many studies have been devoted to developing efficient index compression algorithms [Dean 2009; Navarro et al. 2000; Anh and Moffat 2005; Stepanov et al. 2011]. In particular, many researchers seek to exploit recent hardware features. For example, the SSE instruction sets [Intel 2010] in Intel's processors are collections of Single Instruction Multiple Data (SIMD) instructions introduced with the Pentium 4 in 2001. SSE instructions have accelerated 3D computing [Ma et al. 2002], audio and video processing [Liu et al. 2006], database systems [Willhalm et al. 2013], and other CPU-intensive tasks [Chatterjee et al. 2005]. SSE instruction sets operate on 128-bit registers: they are able to process four 32-bit integers simultaneously. Inspired by this observation, some pioneering studies have incorporated SIMD-based optimization into compression algorithms [Stepanov et al. 2011; Schlegel et al. 2010]. These studies indicate that the speed of index compression can benefit from vectorization. We aim to develop a compression approach to leverage SIMD instructions present in modern processors.

To design a suitable storage layout, we follow earlier work to store separately control patterns from compressed data and adopt a $k$-way vertical data organization [Lemire and Boystov 2015; Schlegel et al. 2010], which makes algorithms easily vectorized by SIMD instructions. Based on such a storage layout, we present a detailed description of the approach and describe strategies to best leverage SIMD-based optimization.

We start from an existing compression algorithm that we wish to vectorize. While the existing algorithm would compress $N$ integers, we compress $4N$ integers to a 4-way data layout. Our approach is sufficiently flexible to accommodate several existing algorithms while providing good performance. We apply the approach to algorithms of four categories covering most of the important practical compression algorithms.

Using our approach, we develop two novel compression algorithms (or algorithm families), i.e., Group-Simple and Group-Scheme. Group-Simple is extended from the traditional Simple algorithms [Anh and Moffat 2005; Anh and Moffat 2006], which can be considered as a word-aligned algorithm; Group-Scheme extends Elias Gamma coding [Elias 1975], which can be considered as a family containing both bit-aligned and byte-aligned variants. Group-Scheme is flexible enough to adapt to different data sets by adjusting two control factors, i.e., compression granularity and length descriptors. We further present the SIMD-based implementations of Group-Simple and Group-Scheme respectively denoted as SIMD-Group-Simple and SIMD-Group-Scheme.

Besides these two families, we also develop Group and vectorized versions of AFOR [Delbru et al. 2012] and PForDelta [Zukowski 2006]. To evaluate the proposed methods, we construct extensive experiments on four public datasets.

The contribution of this paper is summarized as follows:

— Our approach provides a general way to vectorize traditional compression algorithms.

— We develop several novel compression algorithms based on the general compression approach, namely Group-Simple, Group-Scheme, Group-AFOR and Group-PFD. These algorithms cover four major categories of traditional compression algorithms.

— We implement the corresponding vectorized versions of the proposed algorithms, i.e., SIMD-Group-Simple, SIMD-Group-Scheme, SIMD-Group-AFOR and SIMD-Group-PFD. We also examine several important implementation ideas for optimizing the SIMD based algorithms. To the best of our knowledge, it is the first study to implement such a comprehensive coverage of vectorized compression algorithms in a unified approach.

— We conduct extensive experiments on four diverse datasets, including the TREC standard data sets GOV2 and ClueWeb09B, a Wikipedia dataset and a Twitter dataset. Experiments show that our novel SIMD-based algorithms achieve fast decoding speed, competitive encoding speed and compression ratio compared with several strong baselines.

— We integrate the proposed algorithms into an experimental search engine, and examine the performance of different algorithms by the time cost (i.e., query processing performance) and space cost (i.e., index size).

The remainder of the paper is organized as follows. We review the technical background and related studies in Section 2. We present the general compression approach in Section 3. Section 4, Section 5 and Section 6 present the proposed algorithms include Group-Simple, Group-Scheme, Group-AFOR and Group-PFD. We carefully study the encoding format and the compression/decompression procedure together with their corresponding SIMD-based versions. Section 7 presents the experimental results and detailed analysis. Section 8 concludes the paper and discusses possible extensions.

## 2. RELATED WORK

Inverted indexes provide a general way to index various types of documents, whether an article in Wikipedia, a tweet on Twitter, a news article in the New York Times, a status update in Facebook, etc. Although these types of text documents are different in content coverage and presentation format, we can represent all of them as token sequences and create an *inverted index* to facilitate their access. We first introduce the index representations for documents, and then review the existing compression algorithms in four categories.

### 2.1 Background for Inverted Index Compression

#### 2.1.1 Index representations for Web documents

Inverted indexes, as the primary structure of search engines, include two important parts: a dictionary and a collection of posting lists. A term in the dictionary corresponds to a unique posting list. A posting list is composed of a sequence of postings, each posting might correspond to a triplet: <$DocID$, $TF_t$, [$pos_1$, $pos_2$, ..., $pos_{TF}$]>. *DocID* is the document identifier, $TF_t$ is the document frequency of term, and $pos_i$ denotes the document position of the $i$th occurrence of term $t$. We mainly focus on the compression of the frequency and position numbers in the posting lists.

The postings in a posting list are usually sorted in an ascendant order of the value of the *DocID*s. A commonly adopted technique is to perform *d-gap* on posting lists [Manning et al. 2008, p. 96]. Given a strictly increasing positive integer sequence $d_1$, $d_2$, ..., $d_n$, the d-gap operation replaces each integer $d_i$ with $d_i - d_{i-1}$, where $i > 1$. With such a representation, our task is to design ways *to effectively and efficiently compress postings lists in inverted index*. The problem is of general interest: similar techniques are applicable to other fields, such as database compression [Lemke et al. 2011, Raman et al. 2013].

#### 2.1.2 Bit packing techniques

Typically, there are two types of data layout for storing a sequence of encoded integers. *Horizontal layout* stores $n$ encoded integers according to the original order of the integer sequence, while *k-way vertical layout* distributes $n$ consecutive integers to $k$ different groups. Furthermore, the number of bits used to encode an integer in binary format is called *bit width*, while the *effective bit width* denotes the minimum number of bits to encode an integer in binary.

In Figure 1, we present an illustrative example of a 4-way vertical layout. Int1 ~ Int8 denote 8 encoded integers and each four consecutive integers are stored in four different groups respectively.

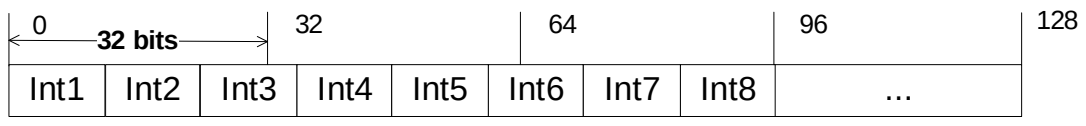

(a) Horizontal layout

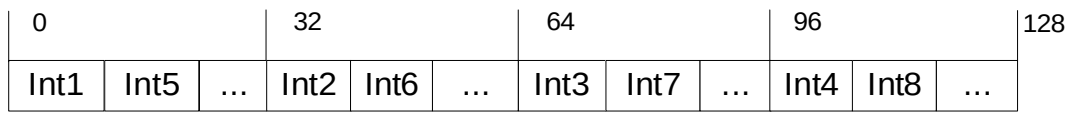

(b) Vertical layout

Fig. 1. Horizontal and vertical data layout.

#### 2.1.3 **Evaluation metrics**

To evaluate the performance of compression algorithms, there are three widely used metrics, namely decoding speed, encoding speed and compression ratio. Decoding/encoding speed measures the processing rate of integers by an algorithm. The compression ratio is defined as the ratio between the uncompressed size and compressed size, and usually expressed as the average number of bits per encoded integer [Manning et al. 2008, p. 87].

### 2.2 **Bit-aligned codes**

In bit-aligned codes, the *bit* is the minimum unit to represent an integer. Such codes can be traced back to Golomb coding [Witten et al. 1999, p. 121] and Rice coding [Rice and Plaunt 1971]. Golomb coding encodes an integer by two parts, i.e., the quotient and the remainder. The quotient is unary encoded, and the remainder is binary encoded. The bit width depends on the divisor *M*, which is commonly set to 0.69 times the average value of all integers. For example, assume that the divisor is 4, we have $14 \div 4 = 3$ (2), thus 14 will be encoded as "0001*10*". Rice coding [Rice and Plaunt 1971] further requires *M* to be a power of two to acceleration the computation. Golomb coding and Rice coding have high compression ratio, but their encoding/decoding speed is low.

Elias Gamma [Elias 1975] encodes an original integer *x* with two parts. The first part is the unary coding of the effective bit width, and the second part is the natural binary representation of *x* without the leading 1. For example, $14=2^3+6$, thus 14 will be encoded as "0001*110*" by Elias Gamma.

Schlegel et al. [Schlegel et al. 2010] proposed a vectorized version of Elias Gamma coding, called *k*-Gamma. *k*-Gamma encodes a sequence of *k* consecutive integers at a time. It first calculates the effective bit width *b* of the maximum integer in this sequence, and represents each integer with *b* bits. Then, the value of *b* is encoded in unary and the low *b* bits of each of *k* integers are encoded in binary. Schlegel et al. adopted the vertical data layout to keep *k* integers word-aligned, and applied SIMD instructions to vectorize for storing and loading the *k* integers. As discussed in Section 5, *k*-Gamma can be viewed as one special variant of our proposed Group-Scheme.

### 2.3 **Byte-aligned codes**

Byte-aligned codes represent an integer in bytes. Variable Byte (VB) encoding [Manning et al. 2008, p. 96] uses bytes to represent a non-negative integer, and the most significant bit of a byte is the continuation bit to indicate whether it is the last byte while the remaining bits store the natural binary representation of the integer.

Group Variable Byte (GVB) [Dean 2009] aggregates the flag bits of a group of integers into a byte called *control byte*. When compressing 32-bit integers, a control byte consists of four 2-bit descriptors, where each descriptor represents the number of bytes needed for an original integer in binary format (00 for 1 byte, 01, for two bytes, 10 for three bytes and 11 for four bytes). GVB encodes and decodes groups of four integers.

There are two ways to code descriptors. If the descriptors are binary coded, the variant is called GVB-Binary. If the descriptors are unary coded, the variant is called GVB-Unary. GVB-Unary includes two variants G8IU and G8CU [Scholer et al. 2002, Stepanov et al. 2011]. Both G8IU and G8CU use 8-byte data areas supported by a 1-byte control pattern. In G8IU, all data areas can be independently decoded, but the last few bytes may be wasted (e.g., when storing integers requiring 3 bytes, only 6 out of 8 bytes can be used). Similarly, the last few bits of the control pattern might be unused. In G8CU, we use all bytes, with integers allowed to overlap between two data areas.

Stepanov et al. exploited SIMD instructions to accelerate the decoding speed of the three variants of GVB [Stepanov et al. 2011]. On x64 processors, integers packed with GVB can be efficiently decoded using the SSSE3 shuffle instructions: *pshufb*. We call Stepanov et al.'s implementations with vectorized decoding as SIMD-GVB-Binary (or SIMD-GVB), SIMD-G8IU and SIMD-G8CU respectively in this paper.

### 2.4 Word-aligned codes

In word-aligned codes, we try to encode as many integers as possible into a 32-bit or 64-bit word. Simple-9 [Anh and Moffat 2005; Anh and Moffat 2006] divides a 32-bit codeword into two parts: a 4-bit selector/pattern and a 28-bit encoded data. It sets up 9 different selectors to instruct the encoding of consecutive integers in the 28-bit data. Zhang et al. proposed Simple-16 [Zhang et al. 2008] improving Simple-9 by extending the number of selectors from 9 to 16.

Anh et al. [Anh and Moffat 2010] used a 64-bit word (a.k.a. Simple-8b and they refer to 32-bit Simple family as Simple-4b). A codeword consists of a 4-bit selector and a 60-bit data.

Previous studies showed that the Simple family has good overall performance with respect to compression/decompression speed and compression ratio [Lemire and Boystov 2015].

### 2.5 Frame based codes

A frame refers to a sequence of integers with the same bit width. This category includes PackedBinary [Anh and Moffat 2010], PForDelta [Zukowski 2006; Zhang et al. 2008; Yan et al. 2009], VSEncoding [Silvestri and Venturini 2010] and AFOR [Delbru et al. 2012]. PackedBinary encodes a frame of integers using the same effective bit width of *b* bits. However, PackedBinary cannot compress well when there are infrequent large integers (called exceptional values).

To deal with exceptional values, Zukowski et al. [Zukowski 2006] proposed PFORDelta (PFD), which separates normal integers from exceptional integers. The normal integers are still encoded with the same bit width, but the exceptional integers are kept in a global exception array and processed separately. Zukowski's implementation does not compress the exceptional values. As a follow-up, Zhang et al. [Zhang et al. 2008] use 8, 16 and 32 bits to store exceptions according to the maximum exceptional values. Yan et al. [Yan et al. 2009] proposed two new variants called NewPFD and OptPFD. They use the same bit width for a frame of 128 integers rather than for all integers. The difference between NewPFD and OptPFD lies in the selection of bit width *b*. NewPFD determines *b* by requiring more than 90% of the integers can be held in *b* bits, while OptPFD determines *b* by optimizing the overall compression ratio.

Lemire and Boystov used SIMD instructions to optimize Packed Binary and PForDelta [Lemire and Boystov 2015]. They proposed a novel vectorized algorithm called SIMD-BP128 for fast decompression. They aggregate 128 successive integers as a frame and use vertical layout to pack them with a unified bit width *b* for the frame. For better compression ratios, they further proposed another new vectorized variant called SIMD-FastPFor, in which they design effective techniques to store the exceptional values.

Packed Binary and PFORDelta adopt a fixed frame length (i.e., the number of integers in a frame) in contrast to approaches using varying frame lengths to improve compression ratio. Silvestri et al. proposed VSEncoding [Silvestri and Venturini 2010], which uses a dynamic programming approach to partition a list of integers into frames. Frame lengths are chosen from the set {1, 2, 4, 8, 12, 16, 32, 64}. Similarly, Delbru et al. proposed AFOR (Adaptive Frame of Reference) [Delbru et al. 2012], which uses only three different frame sizes: {8, 16, 32}.

## 3. A GENERAL SIMD-BASED COMPRESSION APPROACH

In this section, we present a general compression approach designed to incorporate SIMD-based vectorization in integer compression routines. We are motivated by pioneering studies on SIMD-based compression algorithms [Schlegel et al. 2010; Stepanov et al. 2011; Lemire and Boystov 2015]. We borrow and generalize the core ideas of previous SIMD-based algorithms.

For convenience, we first summarize the terminology used throughout the paper in Table I.

Table I. Our terminology.

| Terminology | Explanation |
|---|---|
| Snip | Each primitive unary or binary codeword is referred to as a *snip*. We distinguish between two types of snips: control patterns and data snips. |
| Control pattern | A control pattern is a *snip* that describes how several integers are packed. |
| Control area | The data space storing several control patterns. |
| Vector | A vector denotes 128-bit data. |
| Component | A vector is further divided into four 32-bit data *components*. |
| Data area | The data space containing the data snips where integers are packed. |
| Frame | A frame denotes a sequence of integers. |
| Quadruple | A quadruple denotes four consecutive integers. |
| Scalar | Scalar algorithms use conventional instructions operating on single words. |
| Vectorized | Vectorized algorithms rely on vector instructions operating on several words at once. |

### 3.1 Encoding Formats

The storage layout of a compression algorithm often consists of control patterns and data snips. Data snips represent the encoded integers in binary format, while control patterns code the auxiliary information necessary to interpret the data snips. Many compression algorithms such as VB [Scholer et al. 2002] and Rice [Rice and Plaunt 1971] interleave the control and data snips in a continuous stream. For example, in Simple-9, each 32-bit word contains a control pattern (4 bits) followed by a data snip (28 bits). Yet previous authors have found it convenient to separate control patterns from data snips in distinct space when designing SIMD compression algorithms [Schlegel et al. 2010; Lemire and Boystov 2015]. We adopt this idea. Thus, in creating a vectorized version of a scalar algorithm, we store continuously the data snips in the *data area*. Similarly, we regroup the control patterns in the *control area* separately. Moreover, we pack the data snips using a 4-way vertical data organization.

We consider an existing integer compression algorithm where a control pattern describes a sequence of $N$ consecutive integers. In our approach, we use this same control pattern to describe a sequence of $4N$ consecutive integers. I.e., the control pattern describes $N$ integer quadruples. Each four integers in a quadruple are encoded in the same way, and distributed into the four 32-bit data components of a 128-bit data vector.

As much as possible, our encoding and decoding routines are just a vectorized version of the original scalar routines: if a control pattern specifies that 32 integers are packed using a bit width of 10, then we would pack 128 integers using the same bit width (10). Though simple, this approach is effective.

For algorithms with exceptional values (e.g., *PForDelta*), it is infeasible to directly apply our strategy because we have exceptions in addition to control patterns and data snips. However, it is not difficult to extend our approach to include exceptions stored in a separate location.

### 3.2 SIMD-based Encoding and Decoding

SIMD (Single Instruction, Multiple Data) instructions are widely supported by modern processors. In particular, our SIMD-based algorithms focus on the SSE instructions available on all recent Intel processors [Intel 2010]. These instructions operate on 128-bit registers (called *XMM registers*) making it possible to process four 32-bit integers simultaneously. The main SSE instructions used in our algorithms are:

- **MOVDQA** *dst src*: Copy from 128-bit data source *src* to 128-bit *dst*. *src* and *dst* must be 16-byte aligned, and both cannot be a memory address at the same time (it requires at least one register).
- **MOVDQU** *dst src*: Same as MOVDQA except that *src* and *dst* are allowed to be 16-byte unaligned.
- **PSRLDQ/PSLLDQ** *xmm1 imm8/xmm2*: Regard *xmm1* as an array of four 32-bit integers, and logically shift each integer right/left according the value of immediate *imm8* or *xmm2* register.
- **PAND/POR** *xmm1 xmm2*: execute AND/OR operation on the two 128-bit XMM registers.

It has been noted that a 128-bit data vector can be loaded into 128-bit XMM register, which is particularly useful for the vectorization of the scalar compression algorithms. As discussed in Section 3.1, each four consecutive integers in a quadruple are distributed into four data components by adopting the 4-way vertical layout. More importantly, the four integers in a quadruple are encoded in the same way (e.g., with the same bit width), which makes it feasible to process four 32-bit data components simultaneously with SIMD instructions. For encoding and decoding integers, we are able to vectorize the shift and mask operations for each four integers with SIMD in-structions, which yields a 75% reduction in the number of executed operations. (See Section 4.4.)

### 3.3 Overall organization of the following sections

In Table II, we categorize commonly used (non-SIMD) compression algorithms into four categories. We instantiate the proposed compression approach on several scalar compression algorithms from these four categories. The roadmap of the following sections is listed as follows:

– *Word-aligned*: In Section 4, we propose the *Group-Simple* algorithm, which extends the Simple-9 algorithm.
– *Bit/Byte-aligned*: In Section 5, we propose the *Group-Scheme* family, which originates from the ideas of Elias Gamma and Group Variable Byte algorithms.
– *Frame based*: In Section 6, we propose the *Group-AFOR* and *Group-PFD* based on AFOR and PForDelta respectively.

Table II. Algorithm categorization with the corresponding instantiations in our approach. We mark the instantiated algorithms in bold and present specific modification points to fit into the approach.

| Algorithm category | Scalar algorithms | Instantiations in our approach | Specific modification |
|---|---|---|---|
| Bit-aligned | Golomb<br>Rice<br>**Elias Gamma**<br>*k*-Gamma | Group-Scheme | ● Incorporate different compression granularities and length descriptors into the encoding format. |
| Byte-aligned | Variable Byte<br>**Group Variable Byte** | | |
| Word-aligned | **Simple-9**<br>Simple-16<br>Simple-8b | Group-Simple | ● Provide ten optional controlling patterns and the effective bit width can be up to 32-bits (Simple only supports a maximum bit width of 28 bits). |
| Frame based | PackedBinary<br>**PForDelta**<br>**AFOR**<br>FastPFor | BP128<br>Group-PFor<br>Group-AFOR | ● Apply split selection (AFOR) or bit width selection (Packed Binary) on a quarter of the original integer array. |

## 4. GROUP-SIMPLE ALGORITHM

In this section, we extend the well-known Simple algorithm to a novel algorithm called *Group-Simple*, which uses the general approach in Section 4. Similar to Simple-9/16, Group-Simple still uses four bits to represent a control pattern. The difference is that a control pattern in Group-Simple instructs the compression of 128-bit data rather than 28-bit data. The encoding/decoding operation of a 128-bit data vector can be potentially optimized by SIMD instructions.

### 4.1 Encoding Format and Storage Layout

In this part, we first introduce the storage layout and optional patterns in Group-Simple algorithm.

#### 4.1.1 Storage Layout

In Simple-9/16 algorithms, a 32-bit codeword is divided into two parts: a 4-bit control pattern and 28-bit encoded data. Each control pattern corresponds to a 128-bit data vector, which is further divided into four 32-bit data components.

In Figure 2, we present a schematic diagram for the storage layout in Group-Simple. The major difference between Group-Simple and Simple-9/16 is that control patterns and encoded data are stored separately in physical space to ease vectorization: the control area stores all control patterns, and the data area stores all data components. The data area adopts the 4-way vertical storage layout.

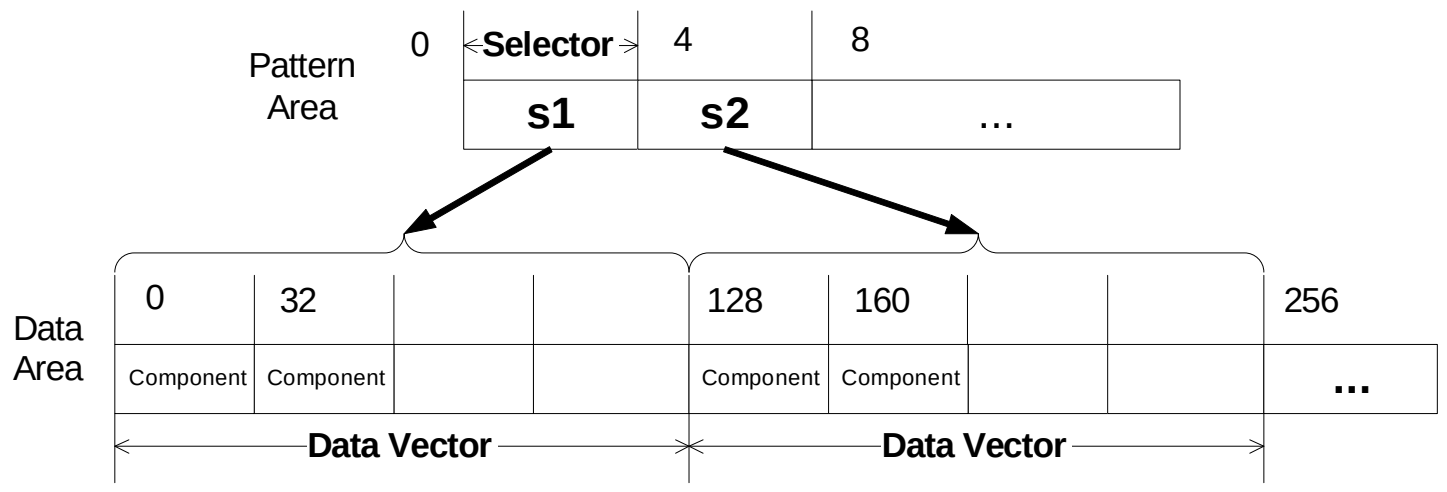

Fig. 2. Overall storage layout in Group-Simple.

#### 4.1.2 Control Patterns

In Group-Simple, we use four bits to represent a control pattern. Each control pattern corresponds to a triplet (*SEL*, *NUM*, *BW*), where *SEL* denotes the selector identifier for the optional patterns, *NUM* denotes the number of integers encoded in a 32-bit data component, and *BW* denotes the bit width of an integer in the data component. Table III shows the ten optional patterns in Group-Simple. With the increase of the selector identifier (*SEL*), the number of encoded integers (*NUM*) in a data component decreases, and the bit width (*BW*) of each encoded integer increases. One advantage of Group-Simple over Simple-9/16 is that the maximum bit width for an encoded integer can be up to 32 bits, which is important for compressing document collections with large *docIDs*.

Table III. Optional patterns in Group-Simple.

| SEL | 0 | 1 | 2 | 3 | 4 | 5 | 6 | 7 | 8 | 9 |
|---|---|---|---|---|---|---|---|---|---|---|
| NUM | 32 | 16 | 10 | 8 | 6 | 5 | 4 | 3 | 2 | 1 |
| BW | 1 | 2 | 3 | 4 | 5 | 6 | 8 | 10 | 16 | 32 |

### 4.2 Encoding Procedure

Our encoding procedure is similar to that of the original Simple-9 algorithm. We determine the control patterns, and store them along the corresponding data snips. In contrast with Simple-9, we store the control patterns and data snips in distinct locations. However our encoding process in preceded by the generation of a temporary buffer called *quad max array* that stores the maxima of quadruples. In what follows, we review the three key points of the encoding procedure in details.

- **Generation of the quad max array.** Group-Simple adopts the 4-way vertical layout, and each four integers from four consecutive components of a vector share the same control pattern. It indicates that each four integers in a quadruple share the same bit width, which is determined by the maximum integer in the quadruple. We refer to this integer as a *quad max integer*. The quad max array *quadmax* stores all the quad max integers. Formally, the quad max numbers are generated taking of the maximum of each four integers in the input array as follows:

$$quadmax[i] = \max(input[4*i], input[4*i+1], input[4*i+2], input[4*i+3]),$$

  where *input* denotes the input array and $i$ is the index variable. The quad max array is built and maintained in RAM, and released once we have generated the control patterns for all integers.

- **Pattern selection algorithm.** Similar to Simple-9/16, a naïve way to select the control patterns is to scan all the integers in a sequence. Here we propose a pattern selection algorithm based on the quad max array, which only needs to process a quarter of the original integers. We present the pattern selection algorithm in Algorithm 1. It requires two input parameters ($i_1$) *MaxArr*, the quad max array and ($i_2$) *L*, the length of *MaxArr*. The algorithm returns with ($o_1$) *ModeArr*, the array of generated selectors and ($o_2$) *TotalModeNum*, the length of *ModeArr*. At each iteration, it examines each of the ten optional patterns in an ascending order of the selector identifiers (*SEL*) and tries to find a pattern to fit the remaining quad max integers as many as possible. With the increase of the selector identifiers, the number of integers in a data vector decreases and the bit width increases. More specially, Line 1~4 are the initialization steps for variables, Line 6~16 are the inner loop for pattern selection, and Line 17~19 are the update steps for the variables. It is worth explaining the inner **for** loop for pattern selector in more details. When an integer has greater effective bit width than the current selector (*BW*), we consider the next pattern with a larger bit width. The loop in Line 14 ends when (1) the number of integers reaches the limit of the current selector, i.e., *NUM* ; (2) we have reached the end of *MaxArr*. We use the shift operation to obtain the largest number with an effective bit width of *b* bits, i.e., the variable *mask*.

**ALGORITHM 1.** The pattern selection algorithm for Group-Simple.

**Input:**

MaxArr, the quad max array of integers with bit widths no larger than 32
L, the length of MaxArr
(NUMi, BWi), the optional patterns in Group-Simple in Table III

**Output:**

*ModeArr,* the array of generated selectors
*TotalModeNum,* the length of *ModeArr*

```
1.  Initialize ModeArr to be an empty array
2.  TotalModeNum ← 0
3.  l ← L  /* get a value copy of L */
4.  j ← 0, k ← 0
5.  while l > 0 do
6.      for i = 0 to 9 do
7.          (n, b) ← (NUM_i, BW_i)
8.          mask ← Power(2,b) - 1
9.          pos ← 0
10.         while pos < min(n,l) AND MaxArr[j+pos] ≤ mask
11.             pos ← pos + 1
12.         end
13.         if pos = n OR pos = l then
14.             exit from for loop
15.         end
16.     end
17.     l ← l – pos, j ← j + pos
18.     ModeArr[k] ← i, k ← k + 1
19.     TotalModeNum ← TotalModeNum + 1;
20. end
```

- **Packing original integers into vectors.** After pattern selection, we examine how to pack a sequence of integers into vectors with the generated control patterns. To encode a single data vector, the algorithm loops *NUM* times. Each time it uses the shift and mask operations to store four integers respectively into four 32-bit data components, where *NUM* is the number of compressed integers in a component. The algorithm will encode 4*NUM* integers into a 128-bit data vector.

Figure 3 presents an illustrative example for the encoding of a sequence of twenty 32-bit integers by using Group-Simple. The quad max integers are marked in bold. To hold the largest integer 64, we need at least a bit width of 6 bits. Therefore, the 5th selector in Table III is selected, which indicates that 5 integers are packed in a 32-bit data component and each encoded integer occupies 6 bits.

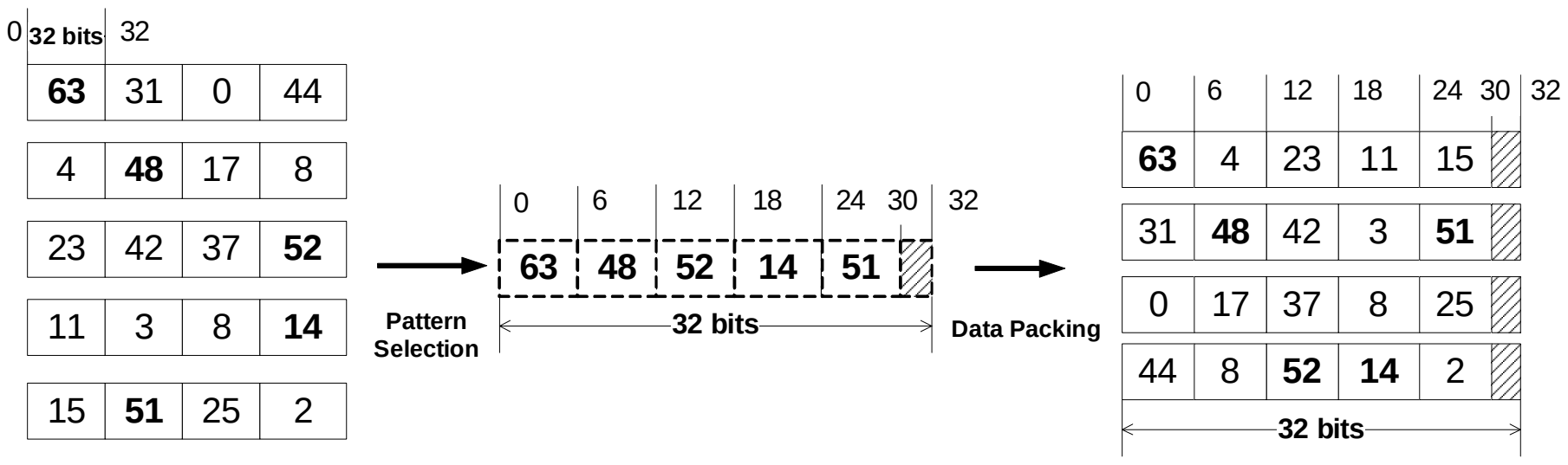

Fig. 3. An example to illustrate the encoding procedure of Group-Simple.

### 4.3 The Decoding Procedure

The key step in decompression is to decode a 128-bit data vector, which contains 4*NUM* encoded integers (*NUM* is the number of integers in a data component corresponding to the selector of the current data vector). In this procedure, the algorithm loops *NUM* times, and in each loop we use the shift and mask operations to decode four integers respectively from four data components.

We present the decoding procedure of Group-Simple as follows:

1) Read the start offset of the data area from the head pointer of *SrcArr*, and locate the start position of the control area, denoted by *ModePos*. The start position of the data area is denoted as *DataPos*.
2) Read four bits from *ModePos* and obtain the current selector.
3) Decode a 128-bit data vector at *DataPos* with the current selector.
4) Move *ModePos* and *DataPos* forward by 4 bits and 128 bits respectively. If *ModePos* does not reach the end, go to step 2.

### 4.4 SIMD-based Implementation and Optimization Techniques

The SIMD-based implementation of Group-Simple is called as *SIMD-Group-Simple*. Once we have transformed the original Simple layout into the format in Figure 2, it is relatively easy to apply SIMD instructions to implement SIMD-Group-Simple: we can vectorize the shift and mask operations and process four integers from four data components simultaneously by using SIMD instructions.

We review two optimization techniques that we put into practice:

- In the encoding procedure, the generation of the quad max array involves conditional statements for value comparison (i.e., identify the maximum value from four integers). The function of the quad max array is to determine the suitable bit widths for encoded integers. To reduce conditional statements, we do not need to identify the exact quad max integers but the *pseudo quad max integers* instead. Following Lemire and Boystov, we use the logical *OR* operations to generate pseudo quad max integers, which may not be equal to the real quad max integer but have the same *effective bit width* [Lemire and Boystov 2015].

- When decoding a single data vector, since both *SEL* and *NUM* have a fixed set of optional values, we use a *SWITCH-CASE* statement with one case for each possible value of the pair (SEL, NUM). For each case we use an optimized routine.

These techniques yield a 20% and 50% improvement at the encoding and decoding speed respectively.

## 5. THE GROUP-SCHEME COMPRESSION ALGORITHM FAMILY

In the previous section, we have presented the instantiation of Simple algorithm with the proposed approach. Inspired by Elias Gamma [Elias 1975] and Group Variable Byte (GVB) [Dean 2009], we present another family of compression algorithms called *Group-Scheme*.

### 5.1 Variants in Group-Scheme family

To better describe the variants in Group-Scheme, we first introduce two terms: compression granularity and length descriptor.

- **Compression granularity** (*CG*) is defined as the minimum unit operated on (or allocated) by a compression algorithm. For example, the compression granularity of Elias Gamma and *k*-Gamma coding is 1 bit, while the compression granularity of Variable Byte encoding is a byte, i.e., 8 bits.

- **Length descriptor** (*LD*) is defined as the minimal number of units necessary to represent (or encode) an integer. For instance, given a compression granularity of 2 bits, the length descriptor of an integer $458_{10}$ ($111001010_2$) is 5 because we need five 2-bit compression units to hold nine bits. We need to represent the length descriptor itself either in binary or unary: e.g., 5 can be represented as “101” in binary or “11110” in unary.

We set up four compression granularities for Group-Scheme: 1, 2, 4 and 8 bits. By combining optional values of compression granularities and the two storage technique for length descriptors (binary and unary), the Group-Scheme family contains eight variants in total as summarized in Table IV.

Table IV. Length descriptors for algorithms in the Group-Scheme family.

| Length Descriptor (LD) | Compression Granularity (CG) | | | |
|---|---|---|---|---|
| | 1 bit | 2 bits | 4 bits | 8 bits |
| Binary | 5 | 4 | 3 | 2 |
| Unary | 1~32 | 1~16 | 1~8 | 1~4 |

When the compression granularity is set to one bit and the length descriptor adopts the complete unary coding, Group-Scheme becomes *k*-Gamma [Schlegel et al. 2010]. In this sense, Group Scheme is a generalization of *k*-Gamma.

### 5.2 Encoding formats and encoding/decoding procedure

#### 5.2.1 Encoding formats

Group-Scheme follows the format of data area described in Section 3, and the major difference lies in the control area, which is composed of several encoded length de-scriptors acting as a control pattern. Based on the coding type (binary/unary), Group-Scheme stores length descriptors (*LD*) in two ways:

(a) *Unary LD:* There are two methods to store unary LDs. If a unary length descriptor can be stored across bytes, we call it *Complete Unary*[1] or *CU* for short. If length descriptors cannot be stored across bytes, we call it *Incomplete Unary* or *IU*. Note that we do not consider the 1-bit and 2-bit CG for incomplete unary LD, since the maximum number of bits needed for a single LD will exceed eight bits with 1-bit and 2-bit CG. Figure 4 shows the examples of IU and CU for the 4-bit and 8-bit compression granularities. Figure 4 illustrates two observations. First, for length descriptors, IU wastes some bits to be byte aligned, while CU does not require byte alignment. Second, for data areas, both IU and CU can store values across words.

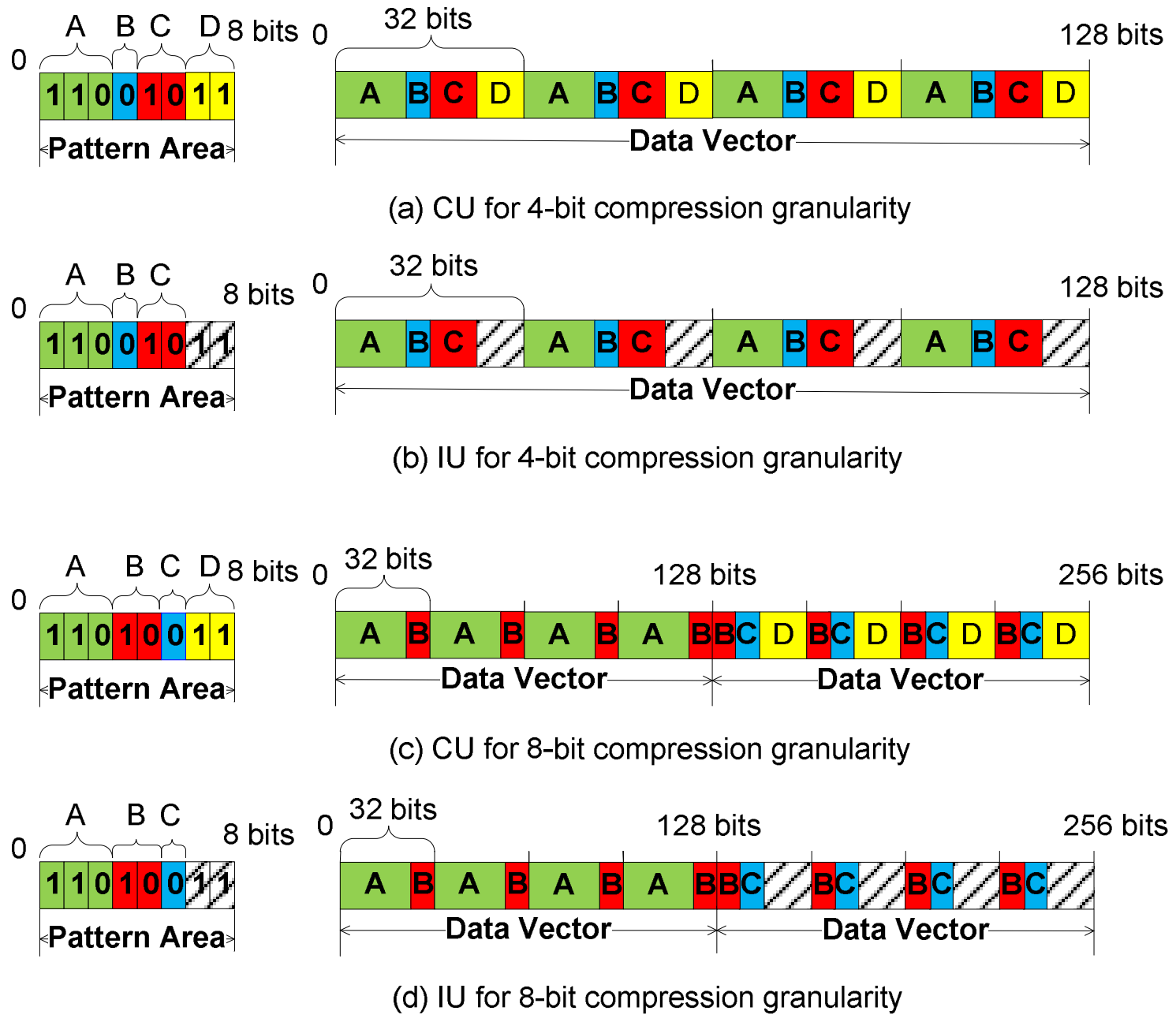

Fig. 4. An example to compare incomplete unary coding (IU) and complete unary coding (CU). Note that there is one data vector in (a) and (b) and two data vectors in (c) and (d).

(b) *Binary LD:* Figure 5 shows four different encoding formats for control area with binary length descriptors. The bit width of a LD is $\lceil \log_2(32/\mathrm{CG}) \rceil = 5 - \lceil \log_2 \mathrm{CG} \rceil$ .We adopt aligned storage for length descriptors at the cost of some wasted bits. The aligned storage is mainly for accelerating and simplifying the encoding/decoding procedure of length descriptors. The alignment depends on the CG. For example, 1-bit CG requires double-byte alignment, while 4-bit CG requires byte alignment. Although binary LDs are word aligned, the data area can store values across words.

[1] Following Stepanov et al., we use the terminologies of complete unary and incomplete unary to discriminate between cross-byte storage and byte-aligned storage [Stepanov et al. 2011].

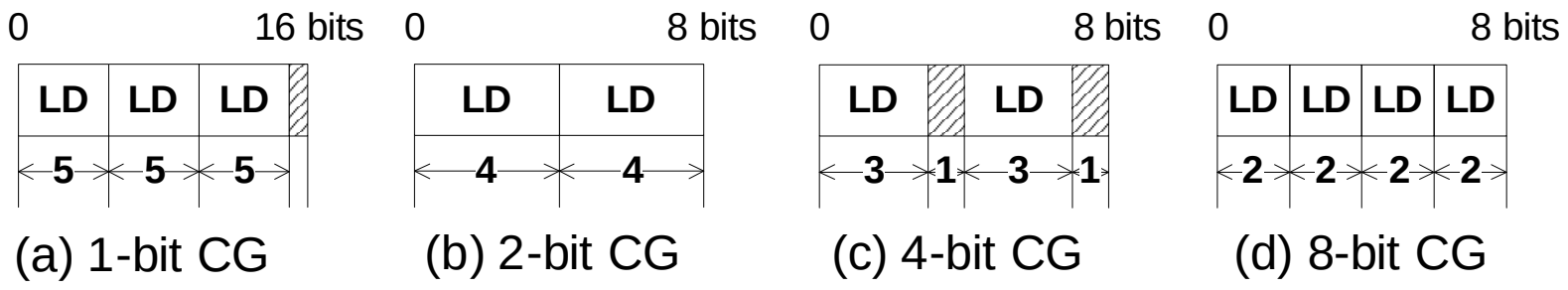

Fig. 5. The corresponding formats of Group-Scheme with different binary length descriptors. The compression granularity and length descriptor are shortened as CG and LD respectively.

#### 5.2.2 Encoding procedure

We first find the maximum integer *quadmax* of a quadruple, and then we calculate the length descriptor as follows

$$\text{ValueOfLD} = \begin{cases} \lceil \lceil log_2(quadmax + 1) \rceil / CG \rceil - 1, & Binary\ LD \\ \lceil \lceil log_2(quadmax + 1) \rceil / CG \rceil, & Unary\ LD \end{cases} \tag{1}$$

The length descriptor (either unary or binary) is stored in the control area. We use the shift and mask operations to encode four integers by taking the low $\lceil \log_2(quadmax + 1)/CG \rceil \times CG$ bits of an integer. Furthermore, these four encoded integers are stored into four different 32-bit data components of a data vector respectively. We update the pointer for the data component and the bit offset within the current component. For an across-word integer, we split it into two parts from high to low by using right-shift and mask operations. The first part is stored in the current data vector and the second one would be stored in the next data vector. The steps are repeated until all the integers are encoded.

#### 5.2.3 Decoding procedure

We first read a length descriptor from the control area and calculate the bit width $BW$ for the encoded integers of data area, where

$$BW = \begin{cases} CG \times (\text{ValueOfLD} + 1), & Binary\ LD \\ CG \times \text{ValueOfLD}, & Unary\ LD \end{cases} \tag{2}$$

We loop four times to extract four $BW$-bit integers respectively from four consecutive 32-bit data components. Then we update the pointer for the data component and the bit offset within the current component. For an across-word integer, we first left-shift the value recovered in the current data vector, and then add it to the value recovered in the next data vector. The steps are repeated until all the integers are decoded.

### 5.3 SIMD-based Implementation and Optimization

Similar to Group-Simple, we can easily implement the SIMD-based version of Group-Scheme, i.e., SIMD-Group-Scheme, which vectorizes the encoding/decoding operations for four consecutive integers in a data vector. In this section, we present several optimization techniques for efficient implementation of Group-Scheme and SIMD-Group-Scheme.

#### 5.3.1 Packed decoding technique for length descriptors

Our experiments empirically showed that the main bottleneck of the decoding procedure lies in recovering the length descriptors. It becomes even worse for unary-coded length descriptors since we have to examine whether the current bit is the end of a length descriptor by using the condition statements. To alleviate this problem, we propose to use the packed decoding technique [Lemire and Boystov 2015] to decode control patterns:

(a) *Unary LD:* At each time, we read a byte instead of a bit and decode all length descriptors in the byte. We compile a lookup table to speed up the decoding process. An 8-bit unary sequence totally has $2^8 = 256$ possibilities. Corresponding to the 256 possibilities, we generate all necessary decoding information and store the information with an array of 256 *STRUCT* elements. Each element contains the following information: (1) the number of encoded integers, (2) the bit width of an encoded integer, and (3) the length of the last consecutive "1"s subsequence without an ending zero (i.e., in the format of "11…1"). For example, a unary sequence "10110011" (left to right) can be decomposed into four parts: "10", "110", "0" and "11". We can obtain three length descriptors: 2, 3 and 1. Note that the last subsequence "11" contains two 1s but does not end with a 0. To deal with such cross-byte length descriptors, we record the incomplete part, and insert it at the beginning of the next 8-bit data.

(b) *Binary LD*: Similar to unary LD, we can use a packed decoding technique by using lookup tables[2]. An extra type of information is needed: the total number of bits actually used in a data component for an 8-bit or 16-bit control sequence. This information can help determine the current pointer for the data component and bit offset within the data component.

Based on our experiments, the *packed decoding technique* for length descriptors can yield about 50% improvement at the decoding speed for most algorithms in Group-Scheme family.

#### 5.3.2 **SIMD-based group unpacking for SIMD-Group-Scheme**

We not only use the packed decoding technique to simultaneously decode several length descriptors, we also use it to decode the data area. For each 8-bit or 16-bit pattern sequence, we have a sequence of SIMD assembly instructions to decode the corresponding integers. This optimization technique is effective to reduce the updating operations of bit offset for vectorized shifting right/left instructions. We describe the implementation details of these assemble functions respectively for unary and binary LD:

(1) *Unary LD:* Assume that there are $4m$ integers to be decoded according to an 8-bit unary pattern data in control area, where $m \geq 1$. If the length descriptor is complete-unary coded, a XMM register will be used to keep the number of unprocessed 1s in the last 8-bit unary sequence we just finished processing in the control area (called XMM1), and another XMM register (called XMM2) is used to keep the corresponding four incomplete integers of the last data vector in data area. The bit offset in the data component is stored in the least significant five bits of XMM3. The bit offset remains in the register and only need to be updated after decoding the $4m$ integers. All the XMM registers would be initialized to zero before decoding a sequence of integers. The steps in the assembly function are as follows:

[Step 1] Load one 128-bit data vector to be decoded into a XMM register
[Step 2] Decode the first four integers in the current vector. Left-shift these four integers by the value in XMM1 and execute vectorized bitwise OR with XMM2. Then the first four integers are decoded and written to memory.

[2] Recall we have four CU for binary LD (See Fig. 5). The size of lookup table for 1-bit CG is $2^{15}$, while the size of the lookup table for 2-bit, 4-bit and 8-bit CG is $2^8$.

[Step 3] Perform the vectorized shift and mask operations to decode the remaining ($4m$-4) integers using lookup tables as in Section 5.3.1, and then write them back to memory.
[Step 4] Write the number of unprocessed unary bits into XMM1, and the corresponding four incomplete integers from data area into XMM2.
[Step 5] Update the bit offset in XMM3.

These steps are designed for complete-unary length descriptors. When using incomplete unary coding, the case becomes simpler: Steps 2 and 4 are not needed again, and a step similar to Step 3 is used to decode $4m$ integers.

(2) *Binary LD:* Similarly, assume that there are $4m$ integers to be decoded according to an 8-bit pattern data. When the length descriptor is coded in binary, the function structure is also simpler because LDs in the control area are word aligned.

In our implementation, the packed decoding technique can yield about 30-100% improvement in the decoding speed for SIMD-Group-Scheme.

### 5.4 Bit Manipulation Instructions

While we use lookup tables to decode unary length descriptors in the Group-Scheme family, we can also decode them with the advanced bit manipulation instructions supported by recent processors. For example, in our experimental servers, two relevant instructions LZCNT (count the number of leading zero bits) and TZCNT (count the number of trailing zero bits) are supported. We can use TZCNT in the encoding procedure and LZCNT in the decoding procedure. The decoding implementation with lookup tables is faster than that of TZCNT with larger compression granularity (i.e., 4 bits and 8 bits) by up to about 10%, but worse with smaller compression granularity (i.e., 1 bit and 2 bits) by up to 30% for sequential algorithms. Indeed, with small compression granularity, we need more bits to encode an integer. In this case, with lookup tables, we can decode fewer length descriptors with a fixed-length pattern sequence. However, for vectorized algorithms, the implementation with lookup tables is general faster (up to 2x) than that of TZCNT. Besides the performance itself, a major benefit of lookup tables is that it is convenient since it applies both for unary and binary descriptors. Based on these considerations, we apply lookup tables to implement all the group-scheme algorithms.

## 6. INSTANTIATION OF THE APPROACH ON FRAME BASED ALGORITHMS

In this section, we review the application of our SIMD-based compression approach to the fourth category of compression algorithms, which splits a sequence of integers into several frames. A *frame* refers to a sequence of integers with the same bit width. We call the instantiated algorithms as Group-AFOR and Group-PFD respectively.

### 6.1 Group-AFOR

The algorithm Group-AFOR is a modification of Adaptive Frame of Reference (AFOR) in [Delbru et al. 2012]. Group-AFOR partitions a sequence of integers into multiple frames of variable lengths. The frame length in AFOR, which is also known as the frame size, is restricted to three values {8,16,32}. To apply our approach, we multiply each frame size by 4: {32,64,128}. The optimal configuration of frame partition and frame lengths is obtained by using an efficient dynamic programing algorithm. The major difference is that we incorporate the quad max array to speed up the encoding. After the partition step, we use the 4-way vertical layout to encode each frame of the original array.

A similar algorithm we considered is VSEncoding algorithm [Silvestri and Venturini 2010]. The major difference between AFOR and VSEncoding lies in the number of optional frame lengths: AFOR provides three lengths while VSEncoding provides five lengths. We implemented both the Group-VSEncoding and SIMD-Group-VSEncoding in our approach. The result of SIMD-Group-VSEncoding was nearly twice as fast as the original VSEncoding algorithm. However, despite the gain in speed, one of our schemes (SIMD-Group-AFOR) offered better speed and better compression ratios. Therefore, we do not discuss VSEncoding further.

### 6.2 Group-PFD

Similar to Group-AFOR, we require the frame size in Group-PFD to be a multiple of four. The major difficulty in wrapping PForDelta is how to process the exceptional integers in the original PForDelta algorithm. We first examine the exceptional entries on the quad max array. Once an exception in the quad max array has been found, we further examine the integers in the corresponding quadruple from the original array.

The detailed encoding procedure for the SIMD-based version of PForDelta (i.e., SIMD-Group-PFD) is described as follows:

[Step 1] Generate the max quad array *MaxArr*.

[Step 2] Run the procedure of calculating the bit width in PForDelta on *MaxArr*: for each frame in *MaxArr*, identify the minimum bit width *b* such that the exception ratio of the frame is below a given threshold $\zeta$ ($0<\zeta<1$). Meanwhile, record the positions of the exceptions in an array *ExOffsetArr*.

[Step 3] According to the bit width of each data vector and *ExOffsetArr*, generate the normal array and exception array based on the original integer sequence: for each exception position in *ExOffsetArr*, we further examine whether each of the integers in the same quadruple is an exception. We update the exception array with the new exceptional positions.

[Step 4] First, we apply SIMD instructions to encode the normal array with the vertical storage layout in the data vector. Then, we encode the exception array by following the Zhang et al. approach [Zhang et al. 2008], which selects the most economical bit width (8, 16, 32).

The decoding procedure is described as follows, which is similar to original PForDelta:

[Step 1] Read the bit width and first exception offset for each data vector.

[Step 2] Use SIMD instructions to decode the normal array.

[Step 3] Use SIMD instructions to decode the exception array and write each exception into the corresponding position of the normal array.

### 6.3 Connections with other frame based compression algorithms

In frame based compression algorithms, typically, a control pattern encodes a sequence of integers (excluding the exceptions), and most of these algorithms can indeed be vectorized in our approach[3], including SIMD-FastPFor, PackedBinary, and SIMD-BP128 [Lemire and Boytsov 2015]. Among these algorithms, SIMD-BP128 achieves the state-of-art decoding speed.

[3] A pattern is extended to encode $4N$ integers, while a pattern encodes $N$ integers.

Lemire and Boytsov's SIMD-BP128 aggregated 128 consecutive integers as a frame and adopted the 4-way vertical layout [Lemire and Boytsov 2015]. SIMD-BP128 packs 128 integers with a fixed bit width, which is the minimum number of bits to hold the largest integer among these 128 integers. SIMD-BP128 can be naturally fit into our compression approach with slight modifications. For SIMD-BP128, the frame size is fixed as 128 and it has a correspondingly low compression ratio (See Section 8).

## 7. EXPERIMENTS

In this section, we first introduce the experimental settings, and then evaluate the proposed compression algorithms with several state-of-the-art algorithms. The evaluation metrics concern the following three aspects respectively: encoding speed, decoding speed and compression ratio. Finally, we examine the performance of different algorithms by the time cost (i.e., query processing performance) and space cost (i.e., index size) in an experimental search engine.

### 7.1 Experimental Settings

(1) **Server profile.** Our experiments are run on a server with an Intel Xeon E5-2620 processor (Sandy Bridge, 2.0GHz, 15MB L3 Cache) with 64GB of DDR3-1333 RAM. The operating system is a 64-bit Linux with kernel version 2.6.32-71. All the compression algorithms are implemented in C++ and complied using GNU GCC 4.6 with the optimization flag "-O3".

(2) **Test collections.** We test the algorithms on several datasets representing different characteristics. The basic statistics of the four datasets are shown in Table V. The first two datasets, TREC GOV2 and TREC Clueweb09B, are released by TREC, which are the standard test collections for the evaluation of the index and retrieval systems. The Wikipedia and Twitter datasets are mainly used to test the algorithm stability on various datasets. Wikipedia[4] is well known for providing formal definitions and knowledge on concepts and entities; Twitter is the most representative microblogging service, and we use the shared dataset by Kwak et al. [Kwak et al. 2010]. For the Twitter collection, we aggregate all the tweets of a user as a document (a.k.a. *user document*), and treat the text in a user profile as the title of the "user document". For these four datasets, we have extracted the title and body fields of each document, and then built an inverted index for each dataset with a search engine system implemented by our group [Shan et al. 2012]. We find that over 90% of d-gap and TF on all four datasets can be represented in 8 bits, which indicates the potential compressibility of these four datasets.

Table V. Basic statistics of the four datasets (1 M=1,000,000) .

| Statistics | GOV2 | ClueWeb09B | WikiPedia | Twitter |
|---|---|---|---|---|
| # Documents | 25M | 50M | 10M | 9M |
| # Terms | 55M | 88M | 46M | 31M |
| # Postings | 5,249M | 11,975M | 1,338M | 1,466M |
| # Tokens | 19,446M | 28,796M | 3,441M | 3,575M |

In what follows, we first present the evaluation experiments on the standard TREC datasets, and then further validate the performance of different algorithms on Wikipedia and Twitter datasets. For TREC datasets, we first randomly selected

[4] http://dumps.wikimedia.org/enwiki/

10,000 queries from the query set of 2005 TREC Terabyte Track[5], and then kept all the unique terms from these selected queries by excluding stopwords. Next we extracted the d-gap sequences and TF sequences from the posting lists of these selected terms, and performed the encoding/decoding operations using different compression algorithms. Finally, we average the performance over all the selected terms on encoding speed, decoding speed and compression ratio as the final results.

(3) **Evaluation metrics.** The results will be presented in the order of decoding speed, encoding speed and compression ratio. To reduce the effect from system disturbance, the final result is the average of ten runs. We use *the number of million integers per second* (*mis*) as the speed metrics, and a larger value indicates better performance. We use *the average bits needed per encoded integer (bits/int)* as the metrics of compression ratio (rate), and a smaller value indicates better performance.

(4) **Methods to compare.** We compared the proposed algorithms with several state-of-the-art compression algorithms. An algorithm named with the prefix "SIMD-" indicate that it has been optimized by SIMD instructions in the encoding/decoding procedure. For both the scalar and vectorized algorithms, we have adopted similar optimization techniques as described in previous sections. We present the comparison algorithms as follows:

- *PForDelta*. PForDelta has been one of the most competitive scalar compression algorithms in terms of encoding/decoding speed. Our implementation of PForDelta adopted the optimization method for compressing the exception values in [Zhang et al. 2008]. The exception ratio is set to 10% (SIMD-Group-PFD also follows the same setting).
- PackedBinary and SIMD-FastPFor [Lemire and Boystov 2015]. We set the frame size to 512 integers and use the open-source implementation provided by Lemire and Boystov.[6]
- Our proposed algorithms include: Group-Simple, Group-Scheme, Group-AFOR, Group-PFD and their corresponding SIMD based versions. As previously mentioned, SIMD-BP128 can be considered as a special variant of the proposed approach, thus we have re-implemented SIMD-BP128 with slight modifications in our approach and take SIMD-BP128 as one of our algorithms, too.

Since Group-Scheme contains many variants, we first examine various combinations of compression granularity and length descriptor, and only keep the variants with better performance in following experiments.

### 7.2 Variant Selection in the Group-Scheme Family

Group-Scheme variants are named in the pattern of "CG-LD", where the compression granularity *CG* can be set to 1/2/4/8 bit(s) and the length descriptor *LD* can be set to B/IU/CU. In particular, $k$-Gamma ($k$=4) can be considered as a special variant of Group-Scheme, i.e., the variant "1-CU". The results of encoding/decoding speed and compression ratio are shown in Figure 6(a) and Figure 6(b) respectively.

[5] http://trec.nist.gov/data/terabyte/05/05.efficiency_topics.gz
[6] http://github.com/lemire/FastPFor

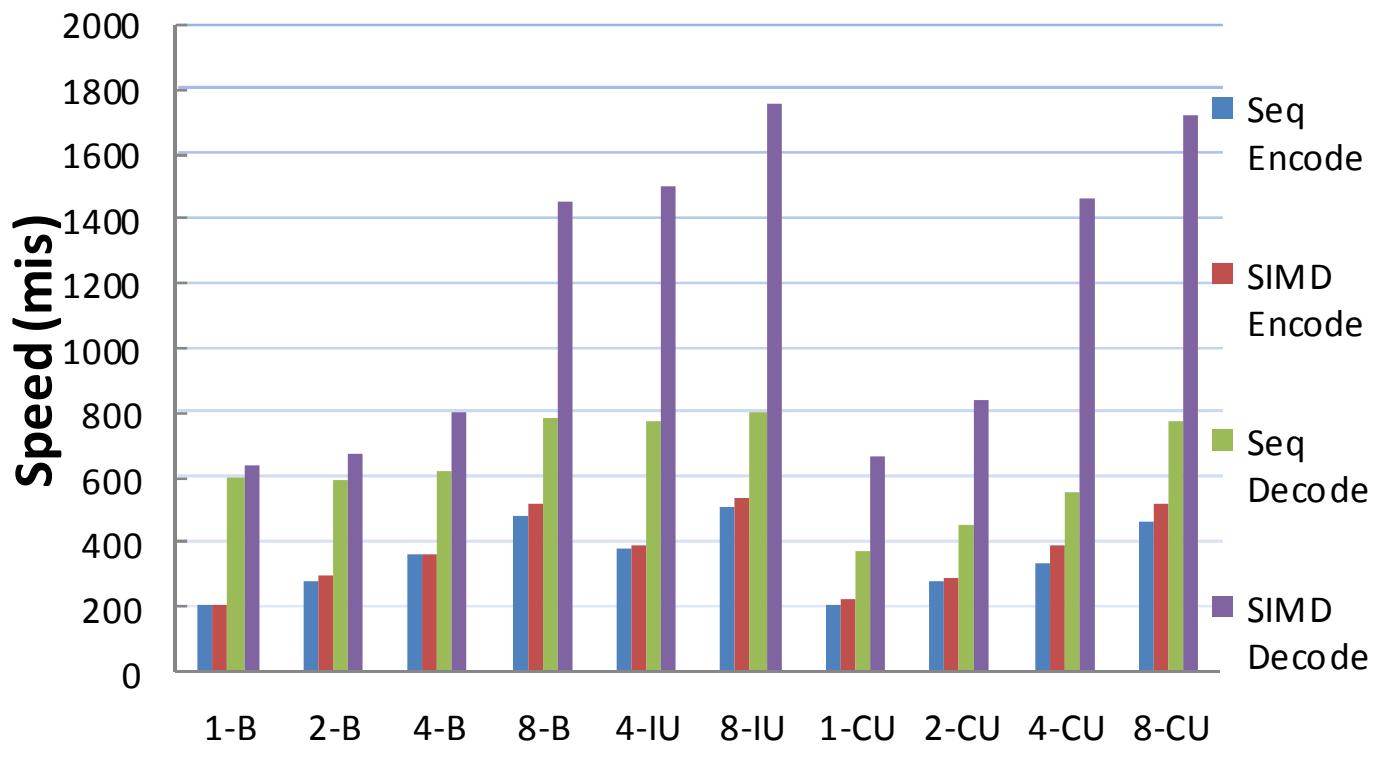

(a) Decoding/Encoding speed

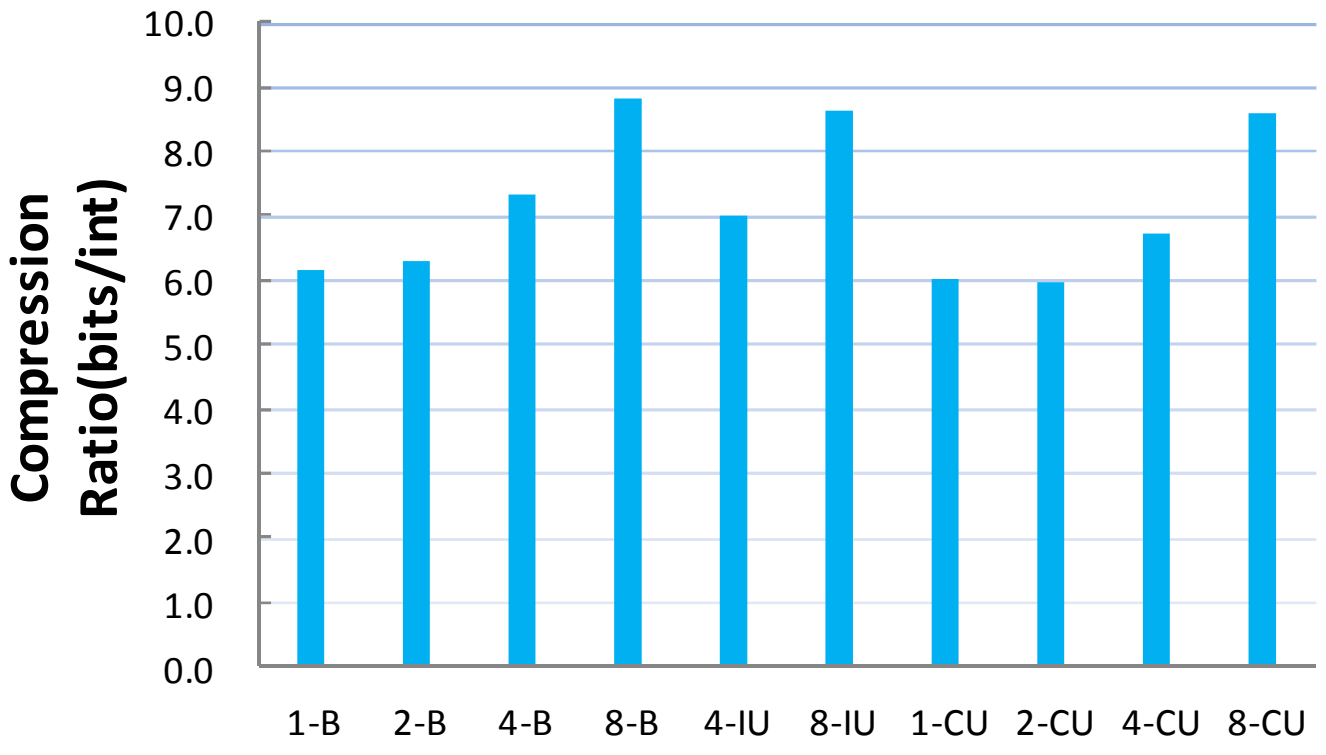

(b) Compression ratio

Fig. 6. Performance comparison on compressing d-gaps of GOV2 within Group-Scheme.

We first analyze the results of decoding speed in Figure 6(a). First, we observe that the SIMD-based algorithms significantly outperform the corresponding non-SIMD algorithms with a large margin varying from 40% to 110%, which indicates the effectiveness of SIMD-based vectorization. For the 1-B, 2-B and 4-B cases, the improvement is small due to SIMD-based vectorization compared to the 8-B case. This is related to the average number of integers we can decode using a pattern sequence. For the 1-B case, a 16-bit pattern sequence can decode 12 integers (6 integers per-byte control data); for the 2-B and 4-B cases, an 8-bit pattern sequence can decode 8 integers. For the 8-B case, an 8-bit pattern sequence can decode 16 integers. For each 8-bit or 16-bit pattern sequence (i.e., 16 bits for 1-B and 8 bits for 2-B, 4-B and 8-B), we pre-generate a sequence of SIMD assembly instructions to decode the corresponding integers. This optimization technique is effective to reduce the updating operations of bit offset for vectorized shifting right/left instructions. Based on our empirical finding, the larger the number of integers a pattern sequence can decode is, the larger improvement the corresponding SIMD-based algorithm is likely to yield. Second, for both the non-SIMD and SIMD-based variants, large compression granularities lead to good performance. The main reason is that large compression granularities correspond to smaller bit widths for LDs, thus the packed decoding technique can process more LDs in an 8-bit pattern sequence. Third, the SIMD-based variants with unary-coded LD (e.g. SIMD-based 4-CU) are faster than those with binary-coded LD (e.g. SIMD-based 4-B) by 20% to 50%. The variant SIMD-based 8-IU achieves the fastest decoding speed, and has a relatively low compression ratio.

For the encoding speed, we have similar observations, but the improvement by incorporating SIMD-based vectorization is small.

We continue to analyze the results of compression ratio in Figure 6(b). With the increase in granularity, the compression ratio becomes worse. Smaller CU stores the encoded integers more compactly, but the corresponding length descriptors take up more space in control area. There is a trade-off between these two types of space cost. In our experiments, the former is the dominant factor for space cost. In addition, the unary-coded LD consistently leads to better compression ratio than binary-coded LD on all compression granularities. For Group-Scheme, variants with larger compression granularity (4 or 8 bits) have faster encoding/decoding speed, and variants with smaller compression granularity (1 or 2 bits) have better compression ratio with competitive encoding/decoding speed. Based on these findings, we select four competitive variants and their corresponding SIMD-based implementation, including Group-Scheme 1-CU (simplified as *GSC-1-CU*), *GSC-8-IU* and SIMD-Group-Scheme 1-CU (simplified as *SIMD-GSC-1-CU*) and *SIMD-GSC-8-IU*.

After variant selection in the Group-Scheme family, we have all the comparison methods ready. We summarize the algorithms to compare and their abbreviations in Table VI.

Table VI. Algorithm Abbreviations. The implementations in our approaches are marked in bold.

| Category | Abbreviation | Algorithms |
|---|---|---|
| Bit-aligned | Rice | Rice |
| | Gamma | Gamma |
| | **GSC-1-CU** | **Group-Scheme-1-CU** |
| | **SIMD-GSC-1-CU** | **SIMD-Group-Scheme-1-CU** |
| Byte-aligned | VarByte | Variable Byte |
| | GVB | Group Variable Byte |
| | G8IU | Group Variable Byte (Unary) |
| | G8CU | Group Variable Byte (Unary) |
| | **GSC-8-IU** | **Group-Scheme-8-IU** |
| | SIMD-GVB | SIMD Group Variable Byte (Binary) |
| | SIMD-G8IU | SIMD Group Variable Byte (Unary) |
| | SIMD-G8CU | SIMD Group Variable Byte (Unary) |
| | **SIMD-GSC-8-IU** | **SIMD-Group-Scheme-8-IU** |
| Word-aligned | Simple-9 | Simple-9 |
| | Simple-16 | Simple-16 |
| | **G-SIM** | **Group-Simple** |
| | **SIMD-G-SIM** | **SIMD-Group-Simple** |
| Frame based | PackedBinary | PackedBinary |
| | PForDelta | PForDelta |
| | AFOR | AFOR |
| | **G-AFOR** | **Group-AFOR** |
| | **G-PFD** | **Group-PForDelta** |
| | SIMD-BP128 | SIMD-BP128 |
| | SIMD-FastPFOR | SIMD-FastPFOR |
| | **SIMD-G-AFOR** | **SIMD-Group-AFOR** |
| | **SIMD-G-PFD** | **SIMD-Group-PForDelata** |

### 7.3 Evaluation on the Compression of Posting Lists

In this section, we evaluate the performance of compressing d-gaps and TF sequences of posting lists. Our Group-AFOR, Group-Simple and Group-PFD algorithms will be abbreviated as G-AFOR, G-SIM and G-PFD respectively in the following tables. We first present the results on GOV2 and ClueWeb09B datasets in Table VII and Table VIII, and then present the results on Wikipedia and Twitter datasets in Table IX. For all tables, we organize the results of algorithms by categories. Note that we consider Group-Scheme as *bit-aligned* since it originates from Elias Gamma coding.

Table VII. Comparison of decoding/encoding speed on GOV2 and ClueWeb09B (mis).

| Category | Algorithm | Decoding Speed | | | | Encoding Speed | | | |
|---|---|---|---|---|---|---|---|---|---|
| | | GOV2 | | ClueWeb09B | | GOV2 | | ClueWeb09B | |
| | | d-gap | TF | d-gap | TF | d-gap | TF | d-gap | TF |
| Bit-aligned | Rice | 67 | 81 | 71 | 84 | 60 | 74 | 61 | 86 |
| | Gamma | 50 | 84 | 58 | 101 | 63 | 85 | 66 | 97 |
| | GSC-1-CU | 376 | 443 | 364 | 475 | 211 | 232 | 205 | 248 |
| | SIMD-GSC-1-CU | 463 | 777 | 534 | 969 | 222 | 243 | 217 | 263 |
| Byte-aligned | VarByte | 538 | 655 | 519 | 671 | 518 | 627 | 495 | 641 |
| | GVB | 514 | 503 | 510 | 506 | 260 | 265 | 253 | 270 |
| | G8IU | 531 | 537 | 508 | 545 | 148 | 121 | 126 | 130 |
| | G8CU | 489 | 510 | 464 | 507 | 146 | 119 | 124 | 124 |
| | GSC-8-IU | 801 | 809 | 740 | 801 | 509 | 516 | 458 | 516 |
| | SIMD-GVB | 823 | 831 | 823 | 823 | 252 | 270 | 247 | 267 |
| | SIMD-G8IU | 1695 | 1719 | 1501 | 1613 | 149 | 125 | 127 | 130 |
| | SIMD-G8CU | 1282 | 1325 | 1243 | 1285 | 145 | 123 | 128 | 127 |
| | SIMD-GSC-8-IU | 1756 | 1772 | 1546 | 1635 | 534 | 552 | 497 | 552 |
| Word-aligned | Simple-9 | 522 | 644 | 506 | 743 | 93 | 114 | 90 | 145 |
| | Simple-16 | 484 | 613 | 500 | 744 | 49 | 61 | 52 | 82 |
| | G-SIM | 943 | 1129 | 1009 | 1266 | 183 | 208 | 178 | 232 |
| | SIMD-G-SIM | 1855 | 1954 | 1577 | 1922 | 242 | 237 | 204 | 265 |
| Frame-based | PackedBinary | 1392 | 1392 | 1252 | 1359 | 244 | 244 | 237 | 244 |
| | PForDelta | 1186 | 1048 | 1028 | 999 | 36 | 42 | 33 | 50 |
| | AFOR | 725 | 579 | 672 | 569 | 247 | 208 | 195 | 211 |
| | G-AFOR | 804 | 756 | 643 | 779 | 242 | 224 | 212 | 223 |
| | G-PFD | 1194 | 1057 | 1020 | 1007 | 89 | 102 | 80 | 120 |
| | SIMD-BP128 | 2273 | 2049 | 1671 | 2155 | 592 | 539 | 318 | 792 |
| | SIMD-FastPFOR | 1912 | 1692 | 1258 | 1405 | 188 | 130 | 139 | 134 |
| | SIMD-G-AFOR | 1976 | 1819 | 1440 | 1673 | 368 | 354 | 366 | 367 |
| | SIMD-G-PFD | 2126 | 1711 | 1695 | 1543 | 220 | 210 | 170 | 192 |

#### 7.3.1 Decoding Speed

In Table VII, nearly all SIMD-based Group algorithms (e.g. SIMD-Group-Simple) outperform the corresponding non-SIMD group algorithms (e.g. Group-Simple) and original scalar algorithms (e.g. Simple). The major observations are listed as follows:

(1) In the bit/byte-aligned category, Rice and Elias Gamma have slow decoding speed. Overall, GVB has relatively better decoding speed, but it is still much lower than our proposed GSC-8-IU. Combining the results in Fig. 6, we can observe that the SIMD-Group-Scheme 1/2/4-CU and SIMD-Group-Scheme 1/2/4-B have slower decoding speed than others of our proposed algorithms, that is mainly because these algorithms have small compression granularity, and thus they tend to spend more time on decoding length descriptors.

(2) In the word-aligned category, Group-Simple is much faster than traditional Simple algorithms (1.5~2.5 times). The major reason is that Group-Simple relates a control pattern to four 32-bit data components while Simple-9/16 relates a control pattern to only a 28-bit data. SIMD-Group-Simple is significantly faster than Group-Simple (1.5~2 times) and traditional Simple algorithms (3~3.8 times).

(3) In the frame based category, SIMD-BP128 has achieved the fastest decoding speed, which is similar to Lemire and Boytsov's finding [Lemire and Boytsov 2015]. As discussed in Section 6.3, SIMD-BP128 can be considered as a special variant of our compression approach. Following SIMD-BP128, SIMD-Group-PFD and SIMD-FastPFor also achieve competitive decoding speed. SIMD-Group-AFOR is two times as fast as Group-AFOR, and Group-AFOR is slightly faster than AFOR.

#### 7.3.2 **Encoding Speed**

Compared to decoding speed, encoding speed is less important than decoding speed since the index is usually built offline. We mainly want to examine whether the incorporation of SIMD instructions can accelerate the encoding procedure. The results are shown in the last four columns of Table VII.

Overall, we observe that SIMD based implementation leads to some improvement in encoding speed, e.g., SIMD-Group-AFOR is faster than Group-AFOR and AFOR by nearly 50%, but the improvement is relatively smaller than that for decoding speed. In addition, our Group algorithms have faster encoding speed: (a) Group-Scheme variants are faster than GVB and Elias Gamma by 3~4 times; (b) Group-Simple is faster than Simple-9 (1.5~2 times) and Simple-16 (2.5~4 times). The key reason for improvement lies in the use of the quad max array. We only need to process a quarter of all the integers with the help of the quad max array.

Table VIII. Comparison of compression ratio on GOV2 and ClueWeb09B (bits per integer).

| Category | Algorithm | GOV2 | | ClueWeb09B | |
|---|---|---|---|---|---|
| | | d-gap | TF | d-gap | TF |
| Bit-aligned | Rice | 5.0 | 3.1 | 5.1 | 2.4 |
| | Gamma | 6.7 | 2.8 | 4.8 | 2.2 |
| | GSC-1-CU | 6.0 | 3.3 | 4.9 | 2.5 |
| Byte-aligned | VarByte | 8.3 | 8.0 | 8.3 | 8.0 |
| | GVB | 10.1 | 10.0 | 10.2 | 10.0 |
| | G8IU | 9.2 | 9.0 | 9.2 | 9.0 |
| | G8CU | 9.2 | 9.0 | 9.2 | 9.0 |
| | GSC-8-IU | 8.6 | 4.9 | 8.8 | 8.3 |
| Word-aligned | Simple-9 | 6.3 | 4.0 | 5.3 | 3.1 |
| | Simple-16 | 5.9 | 3.7 | 5.0 | 2.9 |
| | G-SIM | 6.5 | 4.8 | 6.2 | 3.8 |
| Frame based | PackedBinary/BP128 | 7.0 | 7.1 | 8.8 | 6.1 |
| | PFORDelta | 5.9 | 4.7 | 6.2 | 4.3 |
| | FastPFor | 5.8 | 3.6 | 5.4 | 2.9 |
| | AFOR | 6.0 | 4.0 | 5.3 | 3.1 |
| | G-AFOR | 6.1 | 4.7 | 6.0 | 3.6 |
| | G-PFD | 6.2 | 5.1 | 6.8 | 4.5 |

#### 7.3.3 **Compression Ratio**

Table VIII shows the comparison of compression ratio. Since the incorporation of SIMD instructions does not affect compression ratio, we only present the results of non-SIMD algorithms. Overall, the proposed group algorithms have relatively lower compression ratio compared to the corresponding original algorithms. The main reason is that bit width is determined by the maximum integer in a group, which may potentially lead to more wasted storage. To achieve better vectorization, our algorithms use extra space but still have competitive compression ratios, e.g., GSC-1-CU is slightly worse than the best baselines Rice and Gamma (in terms of compression ratio) but they are 6~10 times faster than Rice and Gamma.

#### 7.3.4 Results on Wikipedia and Twitter datasets

We continue to evaluate different algorithms on Wikipedia and Twitter datasets, which are used to examine the algorithm stability. Due to space limit, we only report the results on d-gaps, and the results on TFs have similar findings.

Table IX. Performance comparison on d-gaps of Wikipedia and Twitter Datasets.

| Category | Algorithm | Decoding Speed (mis) | | Encoding Speed (mis) | | Compression Ratio (bits per integer) | |
|---|---|---|---|---|---|---|---|
| | | Wiki | Twitter | Wiki | Twitter | Wiki | Twitter |
| Bit-aligned | Rice | 67 | 67 | 60 | 60 | **5.4** | **5.5** |
| | Gamma | 48 | 46 | 61 | 61 | 7.2 | 7.5 |
| | SIMD-GSC-1-CU | 445 | 429 | 218 | 219 | 6.3 | 6.6 |
| Byte-aligned | SIMD-G8IU | 1775 | 1841 | 144 | 153 | 9.4 | 9.2 |
| | SIMD-GSC-8-IU | 1936 | 2028 | 515 | 528 | 9.0 | 8.8 |
| Word-aligned | SIMD-G-SIM | 1809 | 1975 | 230 | 238 | 7.0 | 7.1 |
| Frame based | PFORDelta | 1202 | 1264 | 35 | 34 | 6.5 | 6.4 |
| | SIMD-BP128 | **2003** | **2133** | **449** | **507** | 7.8 | 7.7 |
| | SIMD-FastPFor | 2041 | 2210 | 177 | 195 | 6.2 | 6.3 |
| | SIMD-G-AFOR | 1810 | 1929 | 430 | 439 | 6.6 | 6.7 |
| | SIMD-G-PFD | 2230 | 2434 | 198 | 201 | 6.7 | 6.7 |

In Table IX, we can observe that our proposed algorithms still work well on these two datasets: 1) Overall, our SIMD-based algorithms have faster encoding/decoding speed and slightly worse compression ratio than the corresponding non-SIMD algorithms. 2) Rice has the best compression ratio. 3) SIMD-BP128 has the best decoding speed followed by another two competitive algorithms SIMD-Group-PFD and SIMD-FastPFor. 4) Frame based algorithms have similar compression ratio except that SIMD-BP128 has a lower compression ratio (-15%).

### 7.4 Evaluation on query processing performance

In the previous experiments, we have studied the decoding speed of different algorithms. A more direct comparison is to examine the overall performance of query evaluation with different compression algorithms. The time cost for per query processing typically includes the following steps: loading posting lists from disks to memory, decoding d-gaps, decoding TFs, recovering *DocIDs* based on d-gap, locating documents with skip pointers, scoring the candidate documents and top-*k* documents retrieval by using heap sort. Among these steps, only the first three steps are related to compression algorithms, i.e., *loading posting lists from disks to memory*, *decoding d-gaps*, and *decoding TFs*.

In our experiments, we have found that the cost from the disk IO is large and not stable. Therefore, we follow the method of using a warm cache [Delbru et al. 2012], i.e., the time measurements are made when the part of the index read during query processing is fully loaded in memory. The query processing speed is measured by the query rate, i.e., the number of queries a system can process per second. Furthermore, we execute each query ten times, and take the average of ten runs as the final performance for a query.

Table X. Performance rankings on average query processing rate (the number of queries processed per second).

| *GOV2* | | *ClueWeb* | |
|---|---|---|---|
| **Algorithms** | **Query Rate** | **Algorithms** | **Query Rate** |
| SIMD-BP128 | 168.4 | SIMD-BP128 | 68.2 |
| SIMD-G-PFD | 166.7 | SIMD-G-SIM | 67.9 |
| SIMD-G-SIM | 166.4 | SIMD-G8IU | 66.8 |
| SIMD-FastPFor | 164.2 | SIMD-G-PFD | 66.6 |
| SIMD-G8IU | 162.9 | SIMD-FastPFor | 66.2 |
| SIMD-G-AFOR | 161.8 | SIMD-G-AFOR | 65.1 |
| SIMD-GSC-8-IU | 160.5 | PackedBinary | 63.4 |
| PackedBinary | 156.0 | SIMD-GSC-8-IU | 62.6 |
| SIMD-G8CU | 148.1 | SIMD-G8CU | 60.5 |
| G-SIM | 146.0 | Group-Simple | 59.7 |
| PFORDelta | 143.1 | PFORDelta | 57.1 |
| G-AFOR | 134.4 | VarByte | 52.7 |
| VarByte | 130.7 | Group-AFOR | 52.4 |
| AFOR | 127.1 | SIMD-GVB | 50.7 |
| SIMD-GVB | 126.1 | SIMD-GSC-1-CU | 49.8 |
| Simple-9 | 122.2 | Simple-9 | 49.5 |
| SIMD-GSC-1-CU | 119.3 | AFOR | 49.0 |
| GVB | 118.8 | Simple-16 | 48.9 |
| Simple-16 | 118.2 | GVB | 47.6 |
| G8IU | 117.4 | G8IU | 46.9 |
| G8CU | 107.8 | G8CU | 42.5 |
| Rice | 32.1 | Rice | 13.5 |

We still use the GOV2 datasets and the same TREC query set described in Section 7.1 for evaluation. The query evaluation adopts the DAAT (document-at-a-time) scoring way and the top-$k$ retrieval with $k$ set as 10. We use the Okapi BM25 probabilistic model [Robertson et al. 1999] to measure the relevance between a candidate document and a query. Two types of queries are considered: AND query and OR query. We make use of the skipping lists for AND queries. For OR queries, we considered WAND [Broder et al. 2003; Ding and Suel 2011] and MaxScore [Jonassen et al. 2011; Shan et al. 2012].

To better see the advantage of SIMD-based algorithms, we present the results for AND query processing performance descendingly in Table X. We can observe 1) most of the top ten ranks are occupied by SIMD-based algorithms; 2) the proposed Group compression algorithms with SIMD instructions have outperformed the non-SIMD Group algorithms. 3) The SIMD-BP128 and SIMD-Group-Simple achieve competitive performance. In our experiments, decoding of d-gap and TF roughly takes 15%~35% of the overall time cost, therefore the improvement is less significant than that for decoding speeds in Table VII.

The results on OR queries are similar to what have been observed for AND queries, but the overall performance difference between algorithms is relatively small.

### 7.5 Evaluation on index sizes

In this part, we compare the index sizes of different compression algorithms. Generally, the procedure of index creation includes two major steps:

(1) Segment creation: We gradually add documents to the in-memory inverted index. Once it has reached the limit of segment memory, we flush in-memory segment data to the disk. In this step, several local indices will be generated and the posting lists are not compressed.

(2) Index merging and optimizing. We merge all local indices into a global index and perform index optimization. In particular, we focus on the size of posting lists stored on the disk, which consist of three parts in our experiment: d-gaps, TFs and skip list pointers for posting blocks. The skip distance of skip list pointers is set to 512 postings. Skip lists and dictionary take little space. Therefore the overall index size is determined by the size of encoded posting lists. An algorithm with a better compression ratio will generate a smaller index size. Our proposed SIMD-based algorithms take up at least 128 bits no matter how small the size of posting lists is. Therefore, we adopt the traditional compression algorithms for short posting lists: posting lists with a size less than 64 are compressed by Variable Byte encoding.

Table XI shows the index sizes of different datasets by using different compression algorithms. Among all the algorithms, Rice has yielded the smallest indices on all datasets. The overhead for the index can be roughly divided into two parts, i.e., control patterns and encoded data. Group-based algorithms will save space for control patterns by sharing the pattern in the group but take up more space for the encoded data due to the alignment requirement. The final index size is impacted by this trade-off. For example, our algorithms Group-Simple and Group-AFOR take up a bit more space than their corresponding scalar algorithms, i.e., Simple-9, and AFOR; while our algorithms GSC-1-CU (extended from Elias Gamma) and GSC-8-IU (extended from G8IU) result in smaller index sizes compared to Elias Gamma and G8IU respectively.

Table XI. Comparison of index sizes on four datasets with different compression algorithms (MB).

| Category | Algorithm | GOV2 | Clueweb09B | Wiki | Twitter |
|---|---|---|---|---|---|
| Uncompressed | Uncompressed | 40466 | 92062 | 11994 | 12518 |
| Bit-aligned | Gamma | 9571 | 16285 | 4479 | 4140 |
| | Rice | **7280** | **15120** | **3842** | **3387** |
| | GSC-1-CU | 8604 | 16144 | 4021 | 3755 |
| Byte-aligned | VarByte | 12096 | 26378 | 4917 | 4725 |
| | GVB | 14167 | 31382 | 5418 | 5282 |
| | G8CU | 13691 | 29812 | 5715 | 5356 |
| | G8IU | 13852 | 30149 | 5766 | 5416 |
| | GSC-8-IU | 12883 | 28614 | 5176 | 4988 |
| Word-aligned | Simple-9 | 9078 | 17189 | 4236 | 3943 |
| | Simple-16 | 8650 | 16315 | 4139 | 3826 |
| | G-SIM | 9629 | 19948 | 4316 | 4068 |
| Frame based | PackedBinary | 10180 | 23355 | 4460 | 4202 |
| | PForDelta | 8699 | 19434 | 4095 | 3806 |
| | AFOR | 8071 | 15982 | 3870 | 3576 |
| | G-AFOR | 8963 | 18668 | 4118 | 3856 |

## 8. CONCLUSIONS AND FUTURE WORK

In this paper, we studied the problem of optimizing compression algorithms by using SIMD instruction sets. We generalized the ideas of previous studies for SIMD-based vectorization, and proposed a general approach. Based on this approach, we developed several novel compression algorithms, including Group-Simple, Group-Scheme,

Group-AFOR and Group-PFD, together with their corresponding vectorized implementations. We conducted extensive experiments on four large datasets and found that our algorithms performed consistently well on all the datasets. For example, the proposed SIMD-Group-Simple has competitive query evaluation speed. The proposed Group-Scheme has the flexibility to tune on different compression metrics with two adjustable factors: the variants with larger compression granularity have more competitive encoding/decoding speed, and the variants with smaller compression granularity have more competitive compression ratio. We summarize the major highlights of our work in Table XII.

Table XII. Summary of the highlights of our proposed algorithms.

| Algorithm Categories | Algorithms based on our approach | Highlights |
|---|---|---|
| Bit aligned | Group-Scheme | Flexible control over the trade-off between decoding speed and compression ratio with different compression granularity and length descriptor |
| Byte aligned | | |
| Word aligned | Group-Simple | Competitive query processing speed |
| Frame based | BP-128<br>Group-PFor<br>Group-AFOR | The fastest encoding/decoding speed (SIMD-BP128) and competitive encoding/decoding speed (SIMD-Group-PFD) |

In the future, we will study how to apply SIMD techniques on other Intel architecture, i.e., 256-bit or the coming 512-bit vector registers with AVX instructions, and other architectures including PowerPC and ARM. Furthermore, we will test the algorithms on other domains, such as database and image processing.